\documentclass[useAMS,usenatbib,twocolumn]{mn2e}
\usepackage{amsmath}
\usepackage{graphicx}
\usepackage{color}
 \usepackage{multirow}

\def\simlt{\lower.5ex\hbox{$\; \buildrel < \over \sim \;$}}
\def\simgt{\lower.5ex\hbox{$\; \buildrel > \over \sim \;$}}
\def\simpt{\lower.5ex\hbox{$\; \buildrel \propto \over \sim \;$}}

\def\kms{\mbox{km s$^{-1}$}}

\def\kpc{\mbox{kpc}}
\def\pc{\mbox{pc}}

\def\msun{\mbox{M}_\odot}

\definecolor{mylabelcolor}{rgb}{0.5,1,1}

\title[Small-Scale Structures \& Flux Anomalies]{
  Small-Scale Structures of Dark Matter and Flux Anomalies in Quasar Gravitational Lenses}

\date{\today}

\author[Metcalf \& Amara]{R. Benton Metcalf $^{1,2}$  
and Adam Amara$^3$ \\ 
\\
$^1$ Max Plank Institut f\"ur Astrophysics,  
Karl-Schwarzchild-Str. 1, 85741 Garching, Germany \\
$^2$ Dipartimento di Astronomia, Alma Mater Studiorum Universit\'{a} di Bologna, via Ranzani 1, 40127, Bologna, Italy \\
$^3$ Institute of Astronomy, ETH Z\"{u}rich, Wolfgang-Pauli-Strasse 27, CH-8093, Z\"{u}rich Switzerland }

\begin{document}

\maketitle

\begin{abstract}
We investigate the statistics of flux anomalies in gravitationally lensed QSOs as a function of dark matter halo properties such as substructure content and halo ellipticity. We do this by creating a very large number of simulated lenses with finite source sizes to compare with the data.  After analysing these simulations, our conclusions are: 1) The finite size of the source is important.  The point source approximation commonly used can cause biased results.  2) The widely used ${\rm R_{cusp}}$ statistic is sensitive to halo ellipticity as well as the lens' substructure content. 3) For compact substructure, we find new upper bounds on the amount of substructure from the the fact that no simple single-galaxy lenses have been observed with a single source having more than four well separated images. 4) The frequency of image flux anomalies is largely dependent on the total surface mass density in substructures and the size--mass relation for the substructures, and not on the range of substructure masses. 5) Substructure models with the same size--mass relation produce similar numbers of flux anomalies even when their internal mass profiles are different.  6) The lack of high image multiplicity lenses puts a limit on a combination of the substructures' size--mass relation, surface density and mass.  7) Substructures with shallower mass profiles and/or larger sizes produce less extra images.  8) The constraints that we are able to measure here with current data are roughly consistent with $\Lambda$CDM Nbody simulations.
\end{abstract}


\section{introduction}

The Cold Dark Matter (CDM) model with a cosmological constant ($\Lambda$CDM) has become the standard model of cosmology.  This model is in good agreement with a variety of observational probes of the large scales distribution of matter and galaxies in the Universe and is in general agreement with probes of the distribution of mass in galaxy clusters and in large galaxies.  In the $\Lambda$CDM model, dark matter clumps into halos and galaxies form in the halos.  On small scales, $\Lambda$CDM predicts that dark matter halos exist down to very small masses; the exact lower limit depending on the properties of the CDM particle and its thermal history.  It has long been recognized that the number of observed dwarf galaxies in the local group of galaxies falls well short of the number of predicted halos \citep{1999ApJ...524L..19M,1999ApJ...522...82K,2007astro.ph..3337D,2008MNRAS.391.1685S}.  This is referred to as the {\it substructure problem}.   Either galaxy formation is highly suppressed in small mass halos or $\Lambda$CDM needs to be modified in some way by, for example, changing the properties of the dark matter particle or the initial conditions for the density fluctuation in the Universe.  Warm Dark Matter (WDM) is a popular alternative.  Whether or not these small mass halos exist has been one of the most pressing unanswered question in cosmology for a decade.

 \cite{2001ApJ...563....9M} demonstrated that if small-scale structure exists in the distribution of dark matter it will have a strong effect on the magnifications of quasar images in strong gravitational lenses.  This effect causes the flux ratio between images to disagree with any lens model with a smooth distribution of matter.  These cases are call {\it anomalous flux ratios}.  A particular case had been studied by \cite{1998MNRAS.295..587M} and subsequently it was shown that anomalies are common in quasar lenses \citep{2002ApJ...567L...5M,Dalal2002}.  This work and a number of subsequent studies \cite[see][for a review of the subject]{2010AdAst2010E...9Z} relied on fitting lens models to individual lens systems.  It has not yet been shown clearly what can be causing these anomalies and what cannot be causing them.

In a parallel approach, we and others have tried to simulate the lenses directly from cosmological Nbody simulations to determine if they are consistent with the observed frequency of flux anomalies \citep{astro-ph/0306238,AMCO04,2006MNRAS.366.1529M,2009MNRAS.398.1235X}.  The first study predicted a large number of anomalies, but it may have been strongly affected by shot noise.  The two more recent and higher resolution studies found that the substructure in the Nbody simulations is not sufficient to cause the observed flux anomalies (also the conclusion of \cite{2004ApJ...604L...5M}).  This is largely because of the small number density of substructures near the radii where images form (typically around 10~kpc in projection).  These studies relied on only a few projections of a small number of high resolution halos. It is possible that these results are a statistical fluke or that the observed anomalies are largely caused by dark matter objects along the line of sight but not inside the halo of the primary lens \cite{Metcalf04,M04b}.  Answering the question of whether the Nbody simulations have enough small-scale structure in them to account for the flux ratio anomalies is one of the primary goals of this paper.

 It is very difficult to realistically simulate strong QSO lenses from an Nbody simulation.  The first, and most important, problem is that shot noise from the discrete particles has a strong effect on the image magnifications.  Roughly, the error in the magnification goes as $\delta\mu \sim \mu^2 /\sqrt{N_s}$ where $\mu$ is the magnification and $N_s$ is the number of particles over which the smoothing is done.  Since $\mu$ can be large, 100 or larger in the best cases for detecting substructure, the amount of smoothing needed to obtain an accuracy of even 10\% is very large.  So much smoothing can even smooth out the very substructures one wants to detect.  Because of this \cite{2009MNRAS.398.1235X} replace an Nbody simulation with a simple analytic model fit to an Nbody simulation.  A second problem is that the highest resolution simulations do not contain baryons.  Baryons have a strong effect on the profile of the lens and in some cases dominate the mass within one Einstein radius.  The baryons need to be put in ``by hand''.  A third problem is that the extremely high resolution simulations required provide one, or at best a few, dark matter halos. Variations between halos make their lensing properties and their tendency to produce anomalies very different.  It will be demonstrated in this paper that only very limited conclustions about the CDM model can be drawn from a single simultated lens.

To avoid these problems, we take a different approach in this paper.  We produce a large number of analytic lens models that are meant to reproduce the population of lenses expected in the $\Lambda$CDM model.  We then determine the frequency of flux ratio anomalies in these lenses and compare it to the observed frequency.  We adjust the properties and abundance of the substructures to see what kind of substructure is consistent with observations.  The allowed statistical properties of the substructures are compared with the properties of Nbody halos.

All previous studies, except \cite{AMCO04}, have also suffered from the problem that the sources are treated as infinitely small points.  The magnification of individual images are calculated by taking derivatives of the gravitational force at the position of the image.  It will be shown in this paper, that since the physical size of the quasar radio or mid-infrared emission regions are similar to the sizes of the substructures of interest the point source magnifications are not accurate approximations.  We use a new, high speed lensing code called GLAMER (Gravitational Lensing with Adaptive MEsh Refinement) \citep{MetcalfCode} that is the first one capable of producing a very large number of simulated lenses with finite sources in a reasonable amount of time.  It does this through an adaptive mesh refinement algorithm that will be briefly described in section~\ref{sec:ray-shooting}.

In section~\ref{sec:introduction}, the models and techniques used to create simulated lenses are described.  In section~\ref{sec:results}, the results of those simulations are discussed.  Ways of comparing the results to the available lensing data are presented in section~\ref{sec:comparison-with-data}. The results are compared with the predictions of cosmological Nbody simulations in section~\ref{sec:expect-small-scale}.   A summery and discussion are given in section~\ref{sec:concl--disc}.

\section{lens simulations}
\label{sec:introduction}

Our approach in this paper is to produce a large population of
realistic simulated lenses and then compare their statistical
properties to the observed population of lenses.  To do this, we must
develop a model for the population of gravitational lens that includes the host, galaxy + dark matter halo, and the substructures within the host.  We will not consider the effects of companion galaxies with 
masses roughly equivalent to the primary lens in this paper.

\subsection{Host lens model}
\label{sec:host-lens-model}

There is significant evidence from lensing and X-ray observations that
early-type galaxies have a $r^{-2}$ mass profiles
\citep{2007ApJ...667..176G,2010MNRAS.403.2143H,2010MNRAS.404.1165C,2006ApJ...636..698F}.  In accordance with this finding, we
model the host lenses as Distorted Singular Isothermal Ellipsoids
(DSIE).  The surface mass density for this model is
\begin{align}
\kappa(r,\theta)&  \equiv \frac{\Sigma(r,\theta)}{\Sigma_{\rm crit}} \\
& = \frac{r_{\rm E}}{r} \left[ \frac{1}{\sqrt{\cos^2(\theta) +
      f^2\sin^2(\theta)}} \right. \label{eq:kappa1} \\
& ~~~~~~~~ + 
\left.\frac{1}{2}\sum_n b_n(1-n^2) \cos\left[n(\theta +
  \phi_n)\right] \right] \label{eq:kappa2},
\end{align}
where the Einstein radius is
\begin{equation}
r_{\rm E} = 4\pi  \left(\frac{\sigma}{c}\right)^2 \frac{D_lD_{ls}}{D_s}
\end{equation}
and the critical surface density is
\begin{equation}
\Sigma_{\rm crit} =  \frac{c^2}{4\pi G} \frac{D_s}{ D_lD_{ls}},
\end{equation}
where $D_l$, $D_s$ and $D_{ls}$ are the angular size distance to the lens, to the source and between the lens and the source respectively.
The first part (\ref{eq:kappa1}) is a Singular Isothermal Ellipsoids
whose lensing properties have been extensively studied (see
\cite{1994A&A...284..285K} for example).  The deflection angle and shear caused
by the series in  (\ref{eq:kappa2}) have been worked out by \cite{EW03},
although with different notation.

The perturbations $b_n$ are assumed to be of the same order as the observed
perturbations in the surface brightness profile of of early-type
galaxies.  Typical values for $b_3$ and $b_4$ are two or three
percent, but accurate statistics are not available
\citep{1988A&AS...74..385B,2009ApJS..182..216K}.  We draw random
values from a Gaussian distribution with variance 0.005 for $b_2$ and $b_3$ and 0.01 for $b_4$.  We take $n>4$ terms to be zero.  In the observations, $b_4$ is
usually defined with the orientation of this mode fixed to the same axis as
the axis of the elliptical component to define the ``diskyness'' or
``boxyness'' of the galaxy.  Since the alignment has important effects on
the lensing properties, we relax this requirement somewhat and allow
$\phi_{3,4}$ to vary from the position angle of the elliptical component.  
The misalignment is normally distributed with variance 3 degrees.

We also include background shear and convergence in the model.
\cite{2005tyad.confE..12D} calculated the expected distribution of
$\gamma$ and $\kappa$ in an Nbody simulation at potential lenses.
They found that $\kappa$ and $|\gamma|$
are both roughly lognormally distributed with a variance of $\simeq
0.03$.  We assume this distribution in our model.  Analytic estimates by
\cite{1997ApJ...482..604K} are in agreement with this result, as are
observations \citep{2006ApJ...649..599K}.

The model described above is what will be called the ``standard'' host
model.  To test how sensitive magnification anomalies are to the host model,
we perform a series of tests where the distortions to the lens are increased.  For the ``extra distorted model'', we triple the variance in the distortion modes and decouple their orientation from the orientation of the elliptical component.  For the ``extra shear model'', we triple the variance in the background shear and convergence.

\subsubsection{Distributions of host properties}

Calculating the expected distribution of the lenses'
redshifts, velocity dispersions and ellipticities requires knowing not
only the source luminosity and redshift distributions of lenses and
sources, but also the many selection effects that might be important.
The sample of lenses we wish to compare our results with were
discovered in many different ways and do not have a uniform, well
defined selection criterion.  Instead of trying to model these biases,
we use the distributions of already known lenses when possible.

For the lens and sources redshifts, we use the observed values for the
Castles lenses\footnote{http://www.cfa.harvard.edu/castles/}.  There are 60 lenses with measured source and lens redshift
pairs.  We draw randomly from these sets of redshifts.  The lenses
discussed in section~\ref{sec:comparison-with-data} are a subsample of
these.   

To get a sample of host velocity dispersions, $\sigma$, we use the
velocity dispersions from the SLACS lenses
\citep{2006ApJ...649..599K}.  This sample of 61 lenses is used to
make a cumulative distribution of $\sigma$.  The discrete distribution
is linearly interpolated to get a continuous cumulative distribution
and then this is randomly sampled from.  In the SLACS sample, the measured
velocity dispersion of stars and the velocity dispersion of the
best-fit SIE models have statistically indistinguishable distributions.
We choose to use the best-fit SIE velocity dispersions.  These values range from 160 to $396~\kms$. 

The axis ratios, $f$, are sampled independently from the SLACS lenses in
the same way as the velocity dispersions.  No possible correlations
between the internal structure of the lenses and their redshift are
reproduced in this sampling.  The average of this distribution is $f=0.75$, the standard deviation 0.14 and the range is $0.37<f<0.98$. The SLACS lenses are at relatively low redshift because of their selection criterion, but observations indicate that the internal structure of early-type galaxies do not evolve significantly between $z=1$ and 0 \citep{2005ApJ...621..673T}.  

We consider only four image quasar lenses in this paper, while the
SLACS lenses include two image lenses.  The asymmetry of the lens
changes the area enclosed in the tangential caustic and thus a sample
of four image lenses will tend to have more asymmetric lenses than a
sample that includes all multiple image cases.  To correct for this
bias, we calculate the ratio of the area within the tangential caustic
to the area within the radial caustic (or ``cut'' in the case of a
DSIE).  The number of sources used for the lens is then proportional to
this ratio.  More circular galaxies will have less lenses in the final
sample.  This corrects for the bias in the SLACS lenses relative to
the four image quasars.  From 0 to $\sim 100$, source positions are used 
for each lens model.  This method of using a variable number of sources per lens 
is something of a compromise; ideally one would have a population of lenses that 
reflected the biases and one source per lens, but to do this the caustic structure of each lens 
would need to be calculated and then many of those with a small cross-sections for producing four 
images would be discarded.  This would be computationally inefficient.  A small number of sources per lens means that the population of high cross-section lenses will be better sampled, but if the average number of sources per lens is set too low all the lenses with small cross-sections will have zero sources.  We have set the number of sources per lens so that lenses with zero sources are rare ($\sim$ 1\%).

For each lens model the source centers are chosen to randomly cover a region that encloses the region within the tangential caustic.  Some of these source positions give rise to less than four images (when the source intersects the caustic or is completely outside the caustic) and some give rise to more than four images (when caustics structure is more complicated).  The cases with less than four images are discarded in the analysis that follow.

\subsection{Substructure model}

We wish to construct a substructure model that reflects the
expectations we have from Nbody simulation,
but is relatively simple and has a small number of parameters that can
be varied to measure the agreement or disagreement with $\Lambda$CDM.

Simulations show that the mass fraction in substructure within a projected radius increases roughly
linearly with projected radius
\citep{2008MNRAS.391.1685S,2007ApJ...657..262D,2007astro.ph..3337D}.
With a SIE mass model, this implies that the surface mass density of
substructure is constant at least near the Einstein radius and
interior to it.  This will be assumed in all cases.

The mass function of subhalos in Nbody simulations is found to be a power-law
\begin{equation}
\frac{d n}{dm} \propto m^{-\alpha},
\end{equation}
where $n$ is the number of substructures in a halo.  \cite{2008MNRAS.391.1685S} found that 
$\alpha \simeq 1.9$ up to about $1/10$ of the halo mass without any resolved lower mass limit in halos as a whole.  Transforming mass function into a projected mass function in 2 dimensions is not straightforward because of mass segregation in the host halo.   The projected substructure number density will be denoted $\eta$ and the projected mass function will be $d\eta/dm$.

It found that the substructures of different masses are distributed within host halos in remarkably similar ways except that at each radius the mass function has an upper mass cutoff \citep{2008MNRAS.391.1685S}.  If it were not for this mass cutoff the projected mass function (surface number density) would have the same slope as the the total mass function.  Instead, the projected mass function will become steeper than $\alpha = 1.9$ above some mass scale.  We represent this effect in our model crudely with an upper mass cutoff that is smaller than the one found for the complete mass function
\begin{equation}\label{eq:massfunction}
\frac{d \eta}{dm} \propto \left\{ 
\begin{array}{lcc}
m^{-\alpha} &, & m_{\rm min} < m < m_{\rm max} \\
0  &, & {\rm otherwise}
\end{array}
\right.
\end{equation}
with $\alpha \simeq 1.9$.  This is a crude model that could be improved on in the future.

The maximum mass in the mass function must be a function of host halo size.  A mass
scale for the host can be defined as the mass within a fixed radius
($M \propto \sigma^2$) or the mass within a radius where the average
density reaches a fixed threshold ($M \propto \sigma^3$).  The latter is the
one commonly used to define the mass of a halo in cosmology although
the virial radius is generally larger than the radii over which one would
expect the SIE model to hold.  However, if the concentration of the
halos does not vary greatly within the range of host lenses then the same
scaling would be expected in the inner regions.  Making the maximum
substructure mass a fixed fraction of the host halo mass results in
\begin{equation}\label{eq:massscaling}
m_{\rm max}(\sigma) = M_{\rm max} \left( \frac{\sigma}{\sigma_*} \right)^3.
\end{equation}
The same scaling is assumed for the minimum mass.  $M_{\rm min}$ is used as an adjustable 
parameter to change the mass scale and test the data's consistency with a mass cutoff as 
would be expected in many alternatives theories to CDM.  The normalizing halo is fixes to 
$\sigma_* = 200~\kms$. 

The normalization of the mass function~(\ref{eq:massfunction}) needs to
be set.  To agree with Nbody simulations, the fraction of mass in
substructure at a fixed fraction of the virial radius should be the
same in all halos.   Since $R_{\rm host}\propto \sigma$ and
(\ref{eq:massscaling}) makes average mass scale like $\sigma^3$ the
normalization must scale like $\sigma^{-1}$.  Explicitly the result is
\begin{equation}
\frac{d\eta}{dm} = \eta_* \left(\frac{\sigma_*}{\sigma} \right)
\frac{(1-\alpha)}{\left[m_{\rm max}^{1-\alpha}-m_{\rm
      min}^{1-\alpha}\right]} ~ m^{-\alpha}.
\end{equation}
The parameter $\eta_*$ is then the total surface number density of substructures is a host with $\sigma=\sigma_*$ and is not a function of projected radius.

Although the mass fraction in substructure at a fixed fraction of the
halo radius is the same for all lenses, the same is not true at the
Einstein radius.  Since $r_{\rm E} \propto \sigma^2$, the total surface
density at $r_{\rm E}$ is independent of $\sigma$ for lenses and sources at 
the same redshift, which makes the mass fraction scale as $\sigma^2$ at this radius.  
As a result, we might expect substructure to be more important for larger lenses.

The internal structure of the substructures is, for simplicity, a simple
power-law with a cutoff radius
\begin{equation}
\Sigma_{\rm sub}(r) = \left\{  
\begin{array}{ccc}
\frac{(2-\beta)}{2\pi} \frac{m}{R_{\rm cut}(m,\sigma)^2} \left(
  \frac{R_{\rm cut}(m,\sigma)}{r} \right)^{\beta} & , & r < R_{\rm cut}(m,\sigma) \\
0 & , & r > R_{\rm cut}(m,\sigma).
\end{array}
\right.
\end{equation}
In the classical analytic treatment, the average mass density within the
tidal radius is proportional to the average mass density of the host within
the substructure's orbit \citep{BinneyAndTremaine}.  This implies
$R_{\rm cut}(m,\sigma) \propto m^{1/3}$ if all the substructures are at the same
distance from the center of the host, which we assume.  Since the
mass density at a fixed fraction of the host halo radius is independent of
the host size, it is expected that this relation is independent of the
host size:
\begin{equation}
R_{\rm cut}(m,\sigma) = R_{\rm max} \left( \frac{m}{M_{\rm max}} \right)^{1/3}.
\end{equation}
Here $R_{\rm max}$ is a free parameter describing the size of the most massive substructures.  In a more realistic model, there would be a significant
scatter in the $R_{\rm cut}$-$\sigma$-$m$ relation, but for our purposes this relation is sufficient.  
Using the classical tidal radius, the three dimensional distance from the center of the lens that this cutoff radius corresponds to is
\begin{align}
R_{\rm galactic} = 4.3~\kpc \left( \frac{\sigma}{200~\kms}\right)\nonumber \\ \times \left( \frac{R_{\rm max}}{1~\kpc} \right)^{3/2}  \left( \frac{10^9~\msun}{M_{\rm max}} \right)^{1/2}.
\end{align}
Our fiducial model will have $R_{\rm max} = 0.5~\kpc$ and $M_{\rm max} =10^9~\msun$ so  $R_{\rm max} \simeq 1.5 ~\kpc$ is a representative distance which is, perhaps, optimistically compact.  We will vary $R_{\rm max}$ from $0.25~\kpc$ to $4.0~\kpc$.

It should be noted that the appropriate $R_{\rm galactic}$ for lensing would be significantly smaller than the average $R_{\rm galactic}$ for subhalos in general.  Most subhalos are at large radii ($\simgt 100\kpc$) because there is so much volume at large radii to make up for the lower weighted number density.   Projecting along the line-of-sight weights the inner regions of the halo more.  The difference is an order of magnitude or more.  This means that the substructure that are important for lensing will tend to be denser than the overall population.

In summery, the substructure model has the free parameters $\alpha$,
$\beta$, $M_{\rm max}$, $M_{\rm min}$, $R_{\rm max}$, $\eta_*$ and the normalization host velocity dispersion $\sigma_*$ which we fix at $200~\kms$.  However, in the simulations described in the following $\alpha$ and $M_{\rm max}$ are fixed and the remaining parameters are varied.

\subsection{Ray-shooting}
\label{sec:ray-shooting}

\begin{figure} \rotatebox{90}{
\includegraphics[width=9cm]{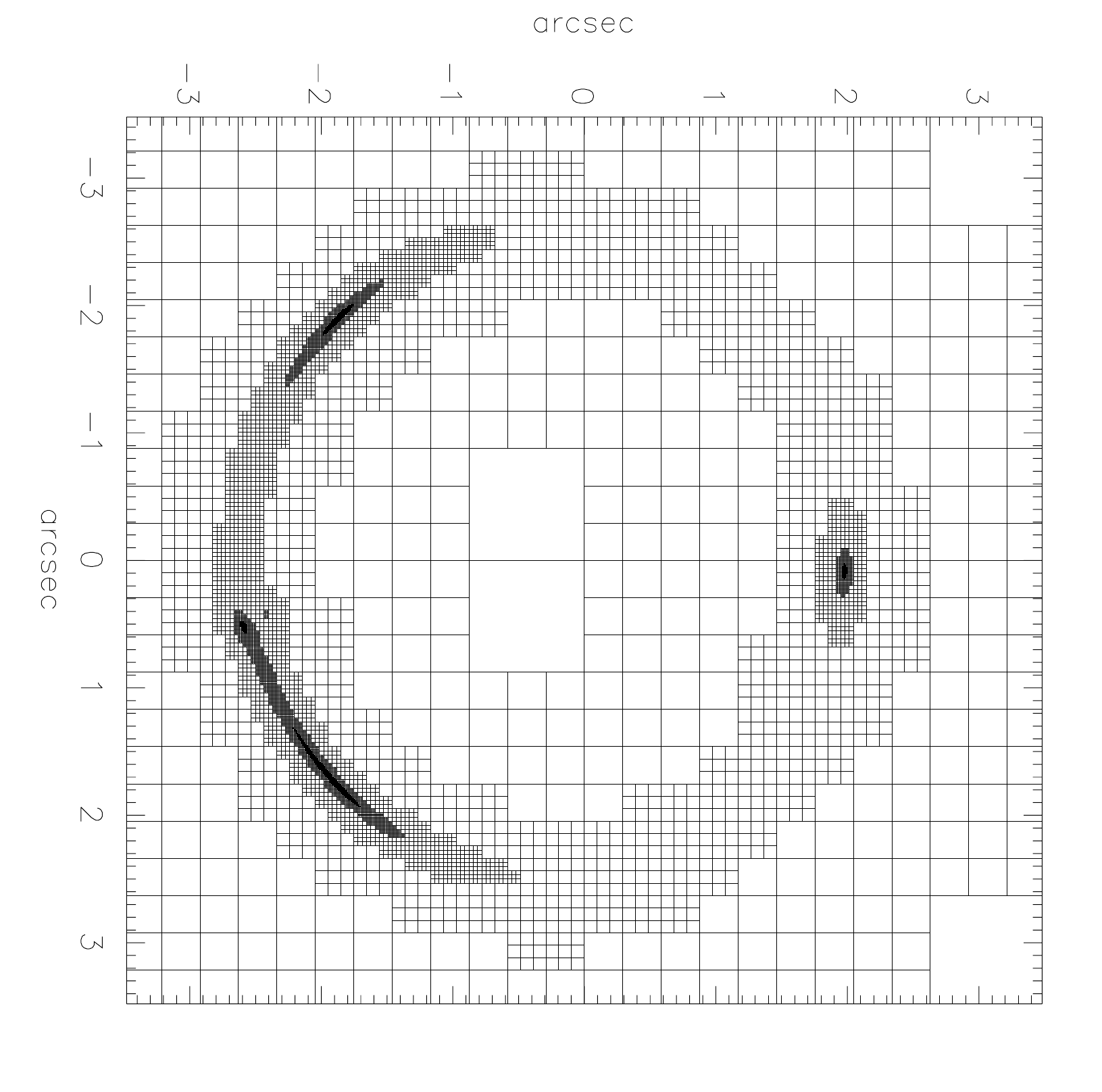}} 
\caption{An example of the refined grid for one particular lens and source position.  The refinements continue below the resolution of this plot.  The deflection angle is calculated once at the center of each grid cell.  There are four images of a 10~pc sources in this case.  At a higher resolution than is visible here, the lower right image breaks into two.} 
\label{fig:grid}
\end{figure}

\begin{figure} \rotatebox{90}{
\includegraphics[width=9cm]{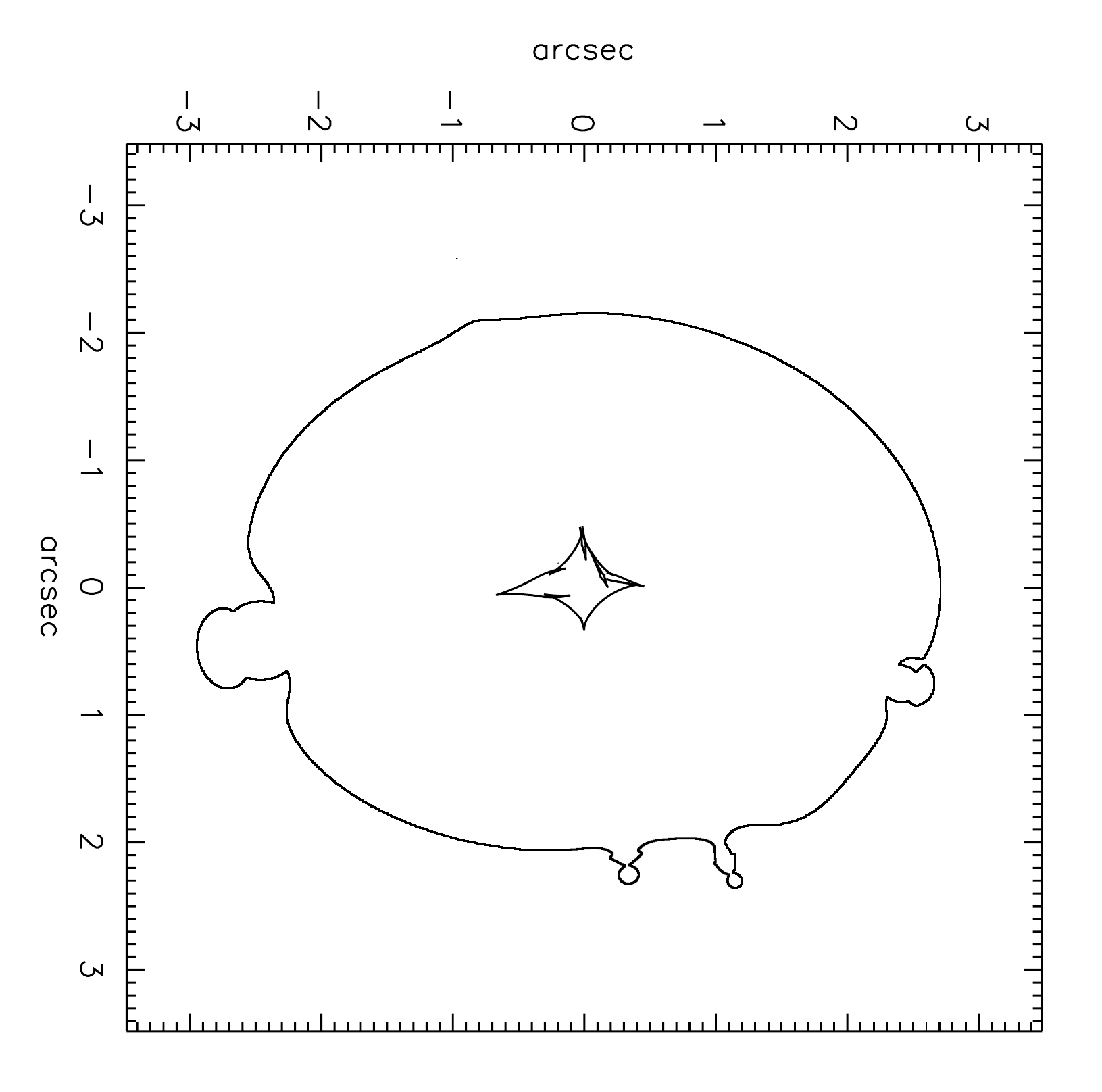}} 
\caption{The critical curve (outer curve) and the caustic for the same lens as in figure~\ref{fig:grid}.  The substructure mass range is $10^7 - 10^9\msun$ with a number density of $\eta_*=0.5~\kpc^{-2}$ and a size scale of $R_{\rm max}=0.5~\kpc$.  In this case the $\sigma = 214~\kms$, $z_{\rm source}=1.34$, $z_{\rm lens}=0.41$ and $f=0.8$.} 
\label{fig:crit_caust}
\end{figure}

\begin{figure} \rotatebox{90}{
\includegraphics[width=9.0cm]{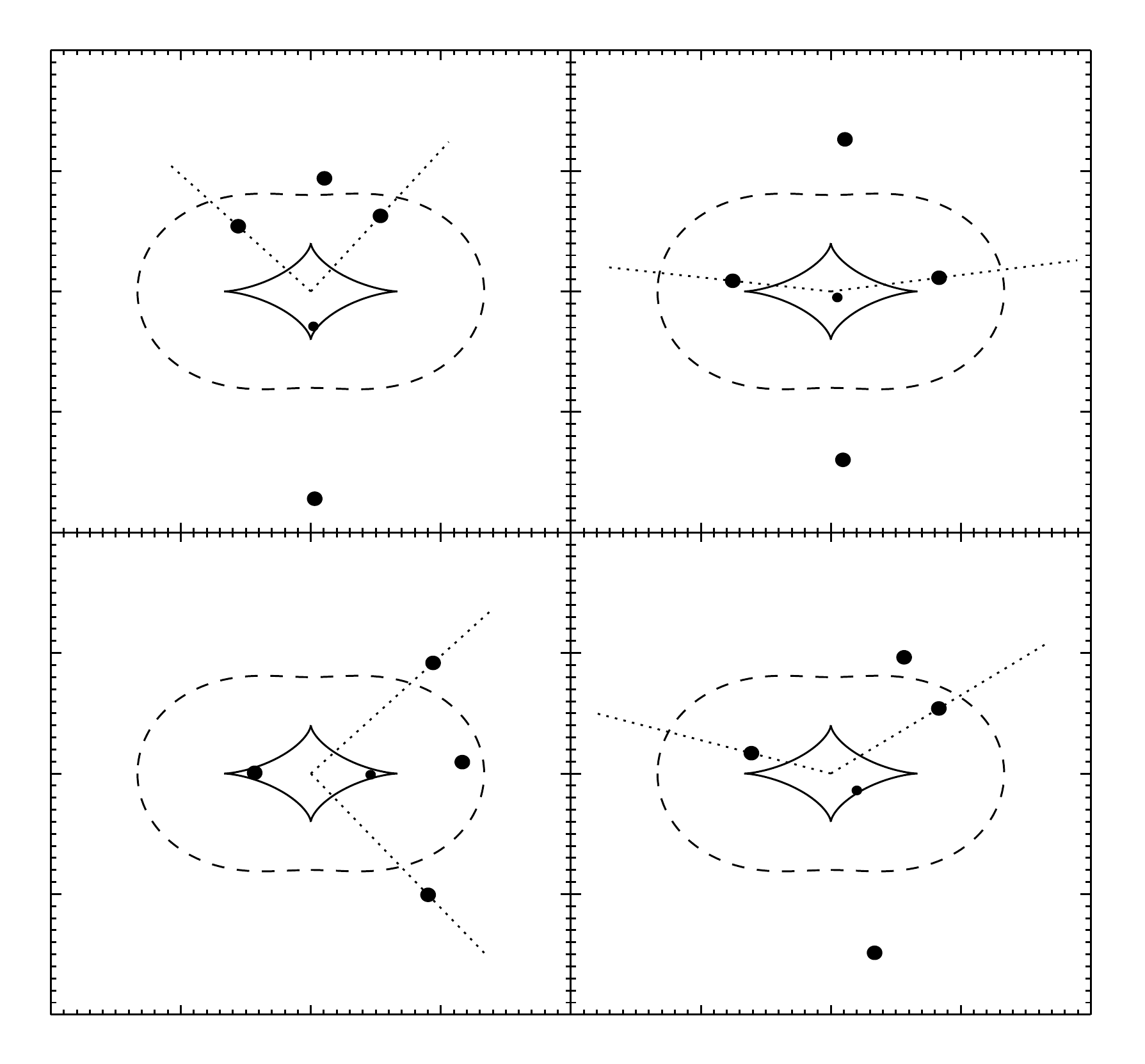}} 
\caption{These diagrams represent the categorization of four image QSO lenses.  The large dots represent the images while the small dot in each panel is the position of the source.  The dashed curve is the critical curve (curve along which the magnification diverges) and the solid curve is the caustic curve (the curve on the source plane that bounds the region in which a source has four images).  The four panels correspond to the four types of lenses.  They are, clockwise from the upper left, an Einstein cross, a fold caustic, a short-axis cusp caustic and a long-axis cusp caustic.  Generally, when the source is near one of the cusps in the caustic, three of the images will be close together. When the source is near the caustic but not near a cusp, two of the images will be close together.  We define the {\it angular separation} between images as the smallest angle between the lines passing through those images and the center of the lens.  The image with the two smallest angular separations to other images is the central image of the {\it image triplet} which includes its neighbors.  It is possible that the triplet is not well defined, but this very seldom happens in practice.  The {\it singlet} image is the remaining image.  The triplet's opening angle, $\Delta\theta$, is the angle between the dotted lines shown in each case.  When $\Delta\theta$ is small, the lens is ``cuspy''.  The categorization of observed lenses into long-axis and short-axis can be made by comparing the distance from the center of the lens to the singlet image, to the distance from the center of the lens to the central image of the triplet.  If the former is larger, it is a short axis case, and if the later is larger it is a long axis case.  In our simulations, this proves to be a very good discriminator.} 
\label{fig:diagram}
\end{figure}

The sources that we wish to use in our simulation have sizes of $\sim 10~\pc$ and the substructures can have similar sizes. Therefore, it is essential that we be able to calculate the magnification of finite size sources.  This requirement has been widely ignored in the literature because it is difficult to map the image of a finite source in a short enough amount of time to make it possible to create the large number of simulated lenses required for this problem.  A new code, GLAMER, has been developed for this and other applications.  
  This code employs a highly optimized adaptive mesh refinement
scheme which allows the shapes of the images and their area to be
calculated rapidly.  (Because of surface brightness conservation, the area of a uniform brightness image is proportional to its magnification.) This allows us to make millions of mock
lenses with a finite size source in a relatively short amount of time.    Figure~\ref{fig:grid} illustrates how the grid is refined to find all the images and their areas.  Figure~\ref{fig:crit_caust} shows the critical curve and caustic structure for one example lens.
For more details on this code, see \cite{MetcalfCode}.

The range of positions in which a substructure will make a significant
change to the magnification of an image depends on the mass of the substructure.  To
optimize calculations, small-mass substructures that are far away from
the lens are omitted from the calculation while more massive substructures further from the lens are included.  To accomplish this, a mass dependent cutoff radius from the center of the lens is used:
\begin{equation}
r_{\rm max}(\sigma,m) = 2 ~r_{\rm E}(\sigma) + R_{\rm cut}(m) + \left( \frac{2\, m\, r_{\rm
    E}(\sigma)}{\pi \Sigma_{\rm crit} \epsilon_{\rm min}} \right)^{1/3}.
\end{equation}
The first two terms ensure that all substructures within two Einstein
radii plus the radius of the substructure are included.  The third
term ensures that any substructure close enough to cause a
perturbation to the lens that is not well approximated as a pure shear
will be included.  The parameter $\epsilon_{\rm min}$ controls how
large the variation in the shear across the Einstein radius are
allowed to be.  We set this parameter to $\epsilon_{\rm min} =
10^{-3}$.  The contribution from substructures or companions
outside this range is considered to be part of the background shear
discussed in section~\ref{sec:host-lens-model} as part of the host lens model.

\begin{table*}
\hspace{-3.5cm}
\caption{Simulation Runs. The top sections show models without substructure, where the source size, distortion and level of external shear is varied. The bottom rows show simulation runs with substructure. $\alpha=1.9$ in all cases.}
\begin{tabular}{c|cllcccccc}
set & host model  & $M_{\rm max}$ ($\msun$) & $M_{\rm min}$ ($\msun$) &
$R_{\rm max}$ (kpc) & $\beta$ & $\eta_*$ ($\kpc^{-2}$) & $R_{\rm source}$ (pc)  & number of simulations \\
\hline
& standard & -  & - & - & - & 0 & 10 & 100,000  \\
& standard & -  & - & - & - & 0 & 1 & 100,000  \\
& extra distorted & - & - & - & - & 0 & 10 & 100,000   \\
& extra shear & - & - & - & - & 0 & 10 & 100,000  \\
& no distortion or shear & - & - & - & - & 0 & 10 & 100,000  \\
\hline
1 & standard & $10^9$ & $10^8$ & 0.5 & 1 & 0.013--0.13 & 10 &  $10^5$ per $\eta_*$ = $1.2\times 10^6$ \\
2 & standard & $10^9$ & $10^7$ & 0.5 & 1 & 0.013--0.40 & 10 & $10^5$ per $\eta_*$ = $3.0\times10^6$ \\
3 & standard & $10^9$ & $10^6$ & 0.5 & 1 & 0.013--0.60 & 10 & $10^5$ per $\eta_*$ = $4.5\times 10^6$  \\
4 & standard & $10^9$ & $10^7$ & 0.5 & 1 & 0.013--0.40 & 1 & $10^5$ per $\eta_*$ = $2.0\times 10^6$  \\
5 & standard & $10^9$ & $10^7$ & 0.25& 1 & 0.013--0.40 & 10 & $10^5$ per $\eta_*$ = $3.0\times 10^6$  \\
6 & standard & $10^9$ & $10^7$ & 1.0 & 1 & 0.013--0.40 & 10 & $10^5$ per $\eta_*$ = $3.0\times 10^6$  \\
7 & standard & $10^9$ & $10^7$ & 0.5 & 0.5 & 0.013--0.40 & 10 & $10^5$ per $\eta_*$ = $3.0\times 10^6$ \\ 
8 & standard & $10^9$ & $10^7$ & 4.0 & 1 & 0.013--0.40 & 10 & $10^5$ per $\eta_*$ = $3.0\times 10^6$  \\ 
9 & standard & $10^{10}$ & $10^7$ & 1.1 & 1 & 0.013--0.41 & 10 & $10^5$ per $\eta_*$ = $3.0\times 10^6$  \\
10 & standard & $10^{10}$ & $10^7$ & 8.6 & 1 & 0.013--0.41 & 10 & $10^5$ per $\eta_*$ = $3.0\times 10^6$ \\
11 & standard & $10^8$ & $10^7$ & 0.23 & 1 & 0.013--0.49 & 10 & $10^5$ per $\eta_*$ = $3.0\times 10^6$ \\
12 & standard & $10^8$ & $10^7$ & 1.8 & 1 & 0.013--0.49 & 10 & $10^5$ per $\eta_*$ = $3.0\times 10^6$ \\
\hline
\end{tabular}
\label{table:simulations}
\end{table*}

For each lens model (host and substructure), the critical curves and caustics are found first.  There are sometimes multiple, disconnected critical curves.  The main tangential caustic is found by requiring its critical curve to be the one that encompasses the most area while also surrounding the center of the lens.  The area within the tangential caustic is calculated and the number of source positions that will be used for that lens is calculated as described in section~\ref{sec:host-lens-model}.   The sources are required to have their centers inside the tangential caustic, but they are otherwise randomly distributed.  Because of the finite source size, some images will be merged and this results in less than four images.

Some lenses have more than the four images that the undistorted host model alone would predict.  Some of these additional images are very small and/or so close to another image that they would not be observed as separate images.  We do a rough initial cut in all cases by merging together any images with centroids that are less than 0.1 arcsec apart, roughly the resolution of the Hubble Space Telescope (HST).  Further discussion of additional images is given in the next section.

Table~\ref{table:simulations} lists the simulation runs that were performed.  They are in batches of 100,000 lenses with fixed substructure parameters.  The first five sets of simulations have no substructure in them and are used to evaluate the importance of distortions to the host lens model and establish a baseline from which to measure the importance of substructure.  The the parameters for the remaining twelve simulation  were chosen to explore the importance of particular substructure properties for lensing.  Set 2 is taken to be a fiducial model.  This is a somewhat arbitrary choice, but we do believe that it is similar to the predictions of Nbody simulations except for the internal profile of the substructures which, as will be shown, has relatively little effect on the lensing properties.  Relative to simulation set 2, set 1 has a higher minimum mass (and average mass), set 3 has a lower minimum mass, set 4 has a smaller source size, set 5 has more compact substructure (a smaller $R_{\rm max}$), sets 6 and 8 have less compact substructure and set 7 has a shallower internal mass profile for the substructures.  In sets 9 and 10, the upper mass cutoff is increased to $10^{10}~\msun$ which is about 10\% of the host's virial mass.  Set 9 has more compact substructures than set 10.   The $R_{\rm max}$ values are set here so that the size--mass relation is the same as in sets 7 and 8.  For eample, a $10^8~\msun$ substructure has the same size in sets 9 and 7.  The rescaling is nessisary because the size--mass relation is normalized at the maximum mass in each model which changes between these models.  In sets 11 and 12, the upper mass cutoff is decreased to $10^{8}~\msun$.  Set 11 has more compact substructures than set 12.  Again the $R_{\rm max}$ values are set to preserve the mass--size relation between sets 7 and 11, and between sets 8 and 12.

The range in surface number density in the simulation sets is meant to span the credible range within a CDM-like model \citep{2007astro.ph..3337D,2008MNRAS.391.1685S}.  In set 3 the number density of substructures is much higher for the same mass density so because of computer time constraints the mass density range for this set does not go as high as in the others although the number density goes higher.  The ranges in $\eta_*$ are were chosen to cover the realistic range in a CDM-like model.  

\section{results}
\label{sec:results}

 \begin{figure*}
\begin{minipage}{16cm}
 \includegraphics[width=16cm]{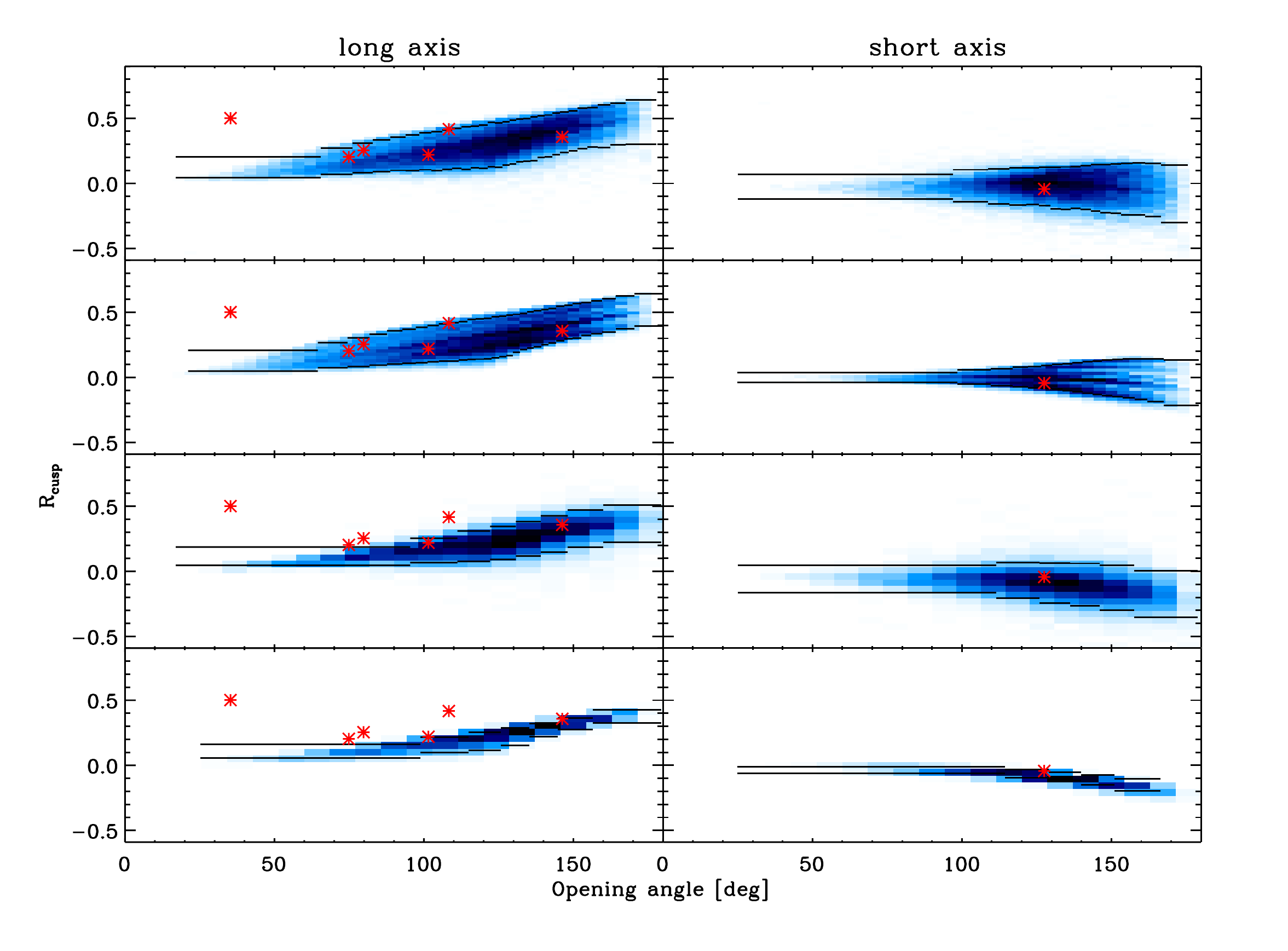} 
 \caption{The distributions of $\Delta\theta$ and $R_{\rm cusp}$ that show the importance of distortion to the elliptical lens model and the ellipticity distribution of the lenses. The blue regions show histograms.  The left column shows the long axis lenses, and the left column shows the short axis lenses.  The top row is for simulation set ``standard'' with $R_{\rm source} = 10$~pc which has random distortions and the full range of ellipticities.  The second row is for the simulation set ``no distortions or shear''  which are pure elliptical models with full range of ellipticities.  The third row is the same as the first, but with all the models with axis-ratio $f < 0.7$ removed.  The fourth row is the same as the second, but with the same axis-ratio cut.  
 The radio and infrared observations are shown as red stars.
It can be clearly seen that the distribution of $R_{\rm cusp}$ is highly dependent on the distribution of lens ellipticities and that most of the observed  $R_{\rm cusp}$ values are not exceptionally high if the full range of ellipticities is considered.   The horizontal lines show where 95\% of the cases are above and 95\% of the cases are below in bins of 3000 simulations.  The observed lenses are shown in red and discussed in section~\ref{sec:comparison-with-data}.}
 \label{fig:Rcusp1}
\end{minipage}
 \end{figure*}

\begin{figure*}
\begin{minipage}{16cm}
 \includegraphics[width=16cm]{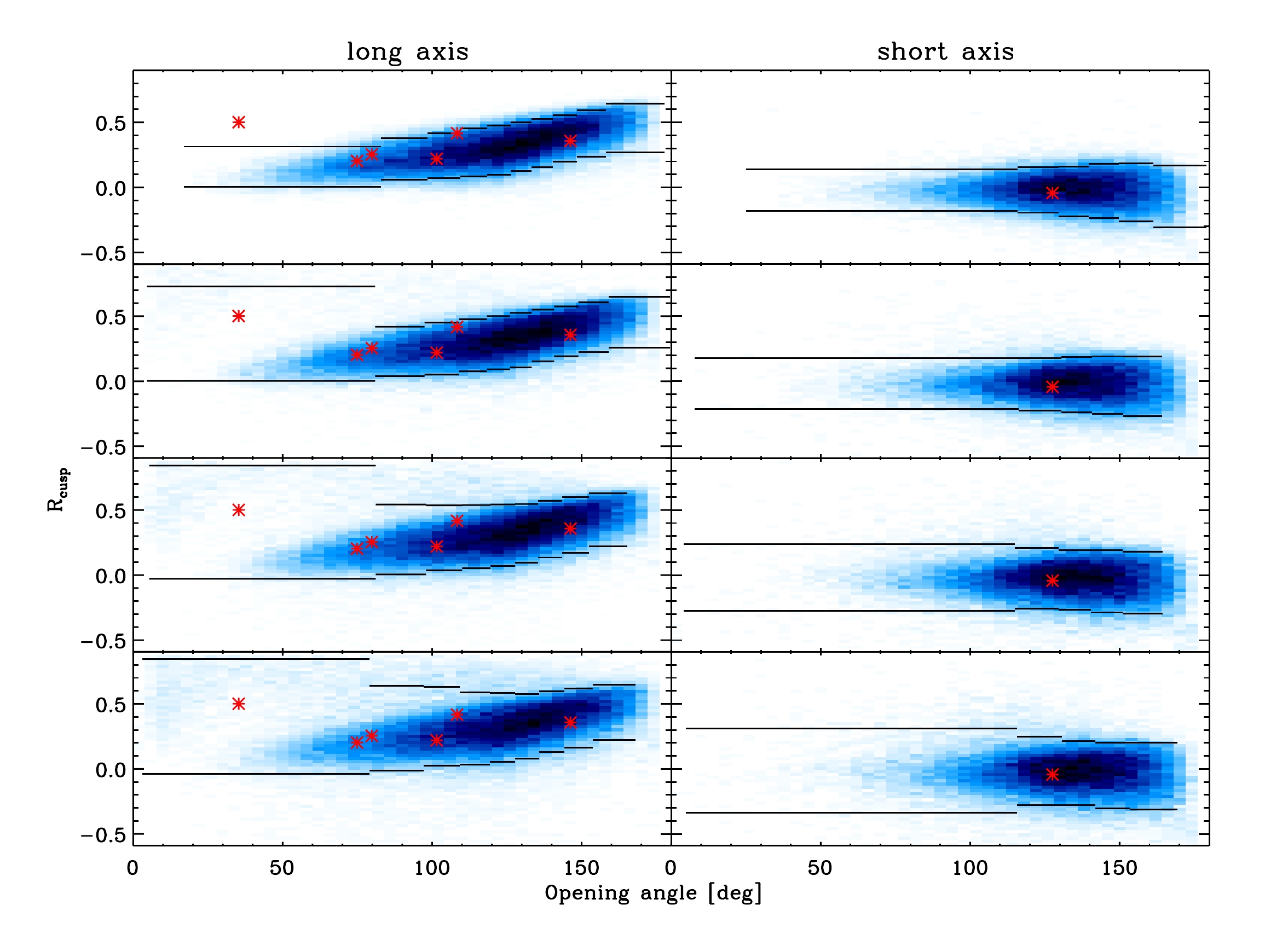} 
 \caption{The distribution of $\Delta\theta$ and $R_{\rm cusp}$ for four of the simulations in set 2. The blue regions show histograms.  The left column shows the long axis lenses, and the right column shows the short axis lenses.  The number density of substructures in each row from top to bottom are $\eta_*= 0$, $0.09$, $0.16$ and $0.27~\kpc^{-2}$.  Random noise of 10\% has been added to represent observational errors.  The horizontal lines show where 95\% of the cases are above and 95\% of the cases are below in bins of 3000 simulations.  These are not exactly the same bins as that are used in calculate the outliers discussed in section~\ref{sec:freq-delt-r_rm}, but are similar.  The radio and infrarad data are shown as red stars.  } 
 \label{fig:Rcusp2}
\end{minipage}
 \end{figure*}

We create several million simulated lenses and save the image positions and magnifications.  We also store the point source magnifications at the centroid of each image and the point source magnification for the point in the image that is closest to the center of the source.  Some of the host lens parameters are also stored.  In this paper, for ease of comparison, we classify the observed and simulated lenses and reduce the position and magnification information to two parameters.  The parameter $\Delta\theta$ is defined in figure~\ref{fig:diagram}.  A small value of $\Delta\theta$ indicates the source is near a cusp in the caustic.  Figure~\ref{fig:diagram} also describes what a long- and short-axis lenses are.   We have found that a good observational way of sorting the lenses into these categories is by comparing
 the angular distance between the center of the lens and the singlet image to the distance between the center of the lens and the central image of the triplet.  If the former is greater, then the lens is a short-axis lens. Otherwise, it is a long-axis lenses.

The second parameter used to characterize each lens is
\begin{equation}
R_{\rm cusp} \equiv \pm ~\frac{\mu_1-\mu_2+\mu_3}{\mu_1+\mu_2+\mu_3},
\end{equation}
where ``$+$'' is for long-axis lenses and ``$-$'' for short-axis lenses.  The magnifications for the images in the triplet are $\mu_1$, $\mu_2$ and $\mu_3$, with $\mu_2$ being for the central image.  The original motivation for this parameter was that $R_{\rm cusp} \rightarrow 0$ asymptotically as a point source approaches a cusp in the caustic \citep{1992A&A...260....1S}. The $R_{\rm cusp}$ parameter has been widely used because of this model independent prediction.  In practice, $R_{\rm cusp}$ is not constrained to a very small region around zero because of finite source effects and the invalidity of the lowest-order expansion of the lensing equation around the cusp.  And, as will be shown, the distribution of $R_{\rm cusp}$ is not very model independent.

 \begin{figure}
 \includegraphics[width=8.5cm]{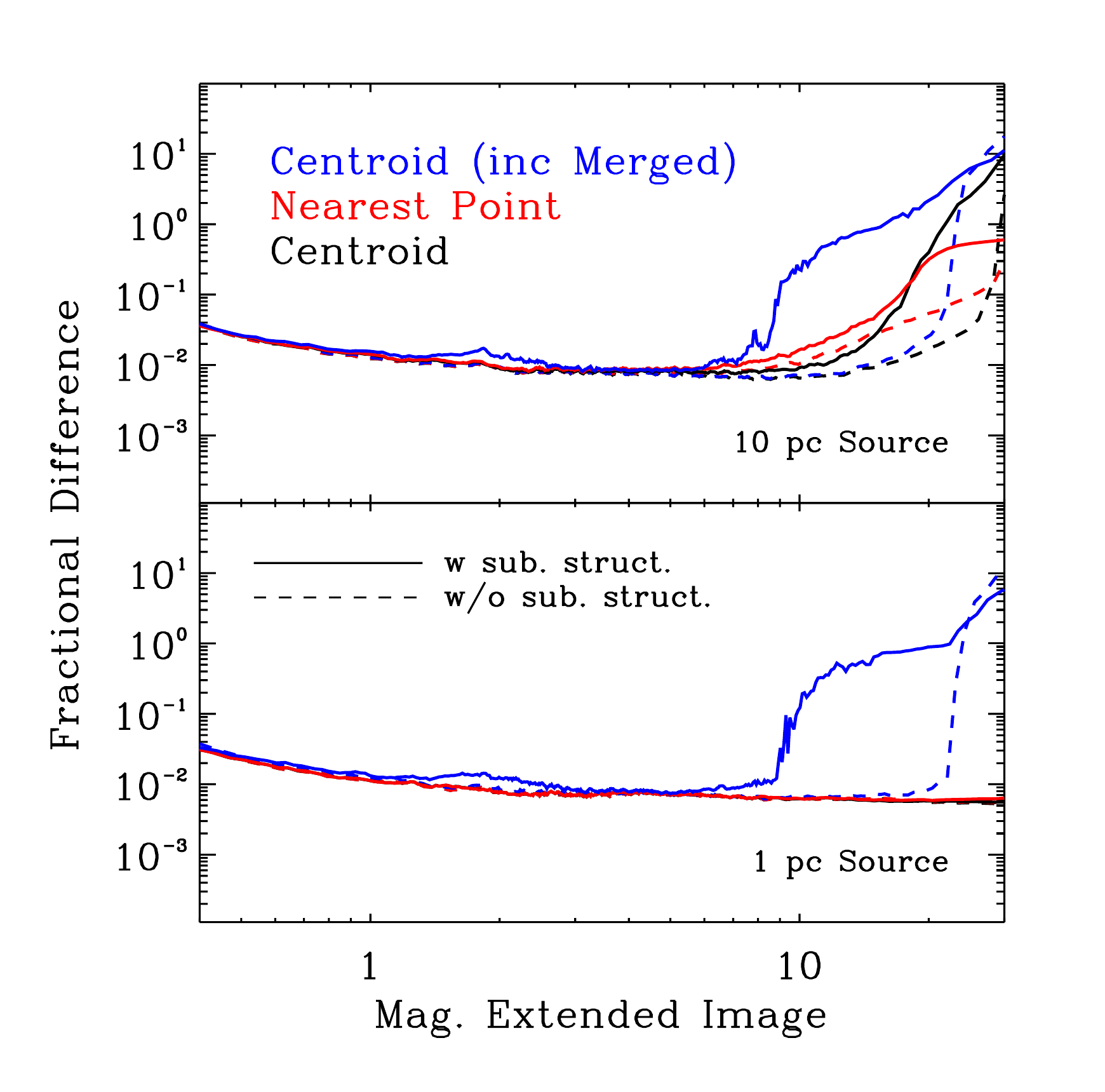} 
 \caption{The range in the fractional error, $(\mu_{\rm point}-\mu_{\rm ext})/\mu_{\rm ext}$. made by using the point source magnification instead of a finite size source.  90\% of the simulations, in running bins of 5000, fall below these curves.  The top panel is for a source with a radius of 10~pc and the bottom panel is for a radius of 1~pc.  The dashed curves are for no substructure ($\eta_* = 0$) and the solid curves are for $\eta_* = 0.2~\kpc^{-2}$ all from simulation set 2.  All the images of all the four-image systems are used. The black curves are for the point magnification calculated at the centroid of the image not including the images that were merged by the $0.1$~arcsec merger requirement.  The blue curves are the same but including the merged cases.  The red curves are for the point source magnification calculated at the grid point in the image that is closest to the center of the source.  The errors for the small source are typically at the 1\% level over a wide range of magnifications, when compared to nearest point estimates, showing good convergence on the level of the numerical noise from the GLAMER ray-tracing code.} 
 \label{fig:mag_error}
 \end{figure}

 \begin{figure}
 \includegraphics[width=8.5cm]{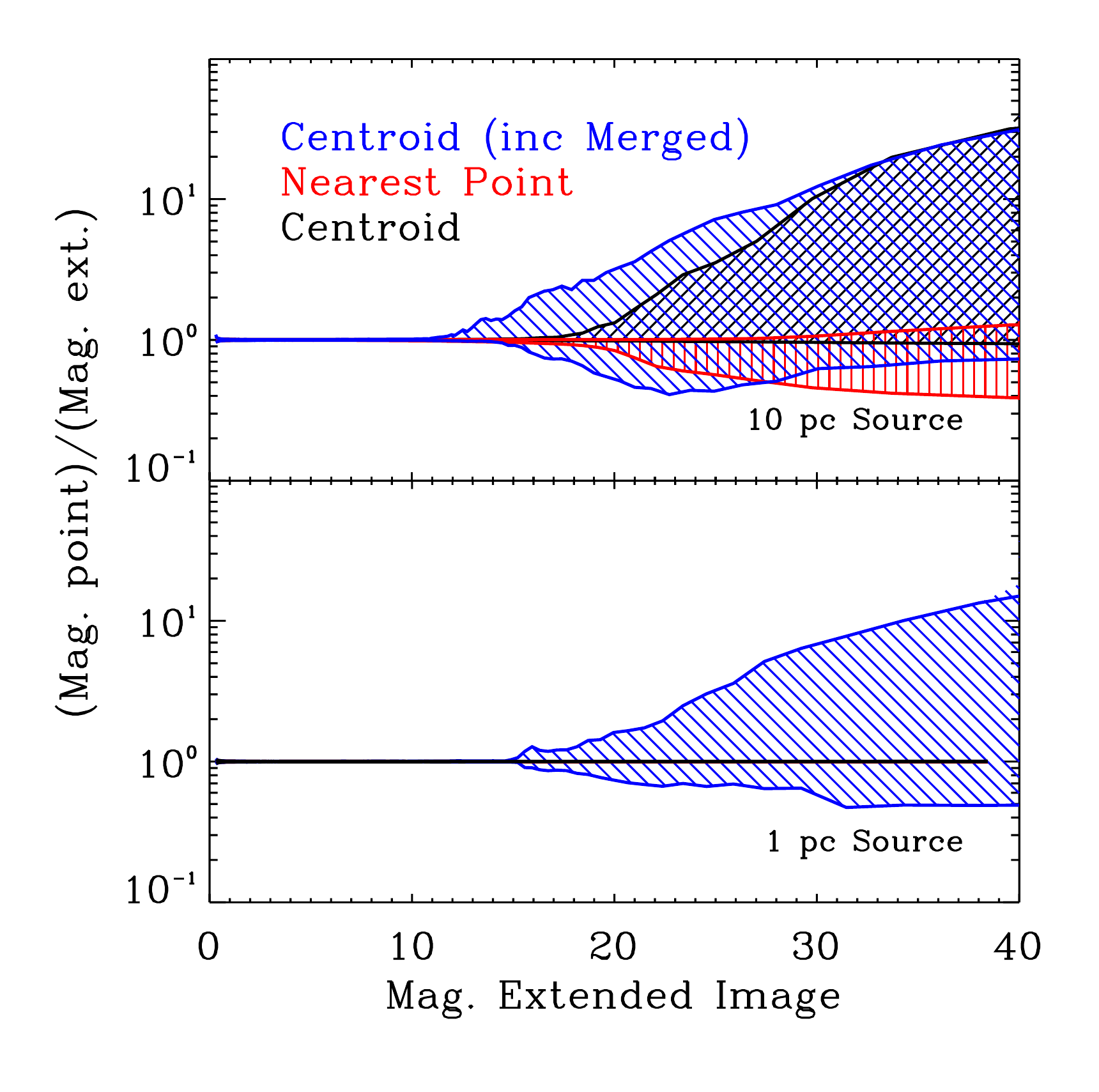} 
 \caption{The ratio of the point source magnification to the finite size magnification.  The shaded regions show where 90\% of the simulations in running bins of 5000 are, 5\% above and 5\% below.  The simulations and color scheme are the same as in figure~\ref{fig:mag_error}.  As in figure~\ref{fig:mag_error}, the upper panel is for a 10~pc source and the lower is for a 1~pc source.  We see that depending on the method used magnification estimates using points will can lead to both random errors and biases (due to the asymmetry of the shaded region) as compared to the extended source calculation, which is closer to the observables. } 
 \label{fig:mag_ratio}
 \end{figure}

Figures~\ref{fig:Rcusp1} and \ref{fig:Rcusp2} shows the distribution of $R_{\rm cusp}$ and $\Delta\theta$ for the sample of simulations listed in the captions.   It can be seen that the simulated lenses occupy a well localized regions in these diagrams when no substructure is present.  Even when substructure is present at the levels investigated, the majority of lenses occupy the same regions with a smaller number of cases spread out in tails to the distribution. 

Figure~\ref{fig:Rcusp1} shows how important the ellipticity of the host lens is to the distribution of $R_{\rm cusp}$ values.  Distortions to the SIE model and background shear do broaden the distribution, but ellipticity has a particularly strong effect.  If only low ellipticity lenses are considered the $R_{\rm cusp}$ values are restricted to a much narrower band.  The sample of lenses is biased toward high ellipticities relative to the general population of lenses because the cross-section for producing four images (the area within the tangential caustic) is increases with increasing ellipticity.  At the same time, Nbody simulations might be biased toward low ellipticity since generally only dynamically well relaxed systems are chosen for very high resolution simulations.  This can explain some of the discrepancies between simulations and observations that have been reported \citep{2010arXiv1007.1599M,2009MNRAS.398.1235X,2006MNRAS.366.1529M}.  This will be further discussed in section~\ref{sec:expect-small-scale}.

Figure~\ref{fig:Rcusp2} is similar to figure~\ref{fig:Rcusp1}, but the effect of substructure on the $\Delta\theta$-$R_{\rm cusp}$ distribution is illustrated. An additional 10\% error on each image's flux is added to conservatively account for typical obervational uncertainties.  Substructure has the effect of producing a papulation of extreme outliers in this distribution. 

Figure~\ref{fig:mag_error} shows the fractional error made in the magnifications when the point source magnification is used.  It can be seen there that the fractional error is small for magnifications less than around 5.  This is confirmation that the numerical errors made by the ray-tracing code are small.  At higher magnifications, larger errors are made when the source is 10~pc.  This is not a numerical effect.  It can also be seen in figure~\ref{fig:mag_error} that substructure causes the errors made by using the point magnification to increase when the source size is 10~pc, but less so when the source size is 1~pc.  This is in agreement with expectations because the source size of 10~pc is closer to the characteristic scale of the substructures.

Figure~\ref{fig:mag_ratio} shows the ratio between the point source magnifications and the finite source magnifications.  Again, it can be seen that numerical errors are not playing a large part.  It is evident that the point source magnifications are not evenly distributed around the finite source magnifications.  Centroid point source magnifications tend to overestimate the real magnification; in some cases by a large factor.  This is the magnification that would be calculated when fitting a lens model to an observed lens.  In the simulation, the centroid is calculated by doing a flux weighted average over the pixels on the simulation grid.
The nearest point magnification is much less biased and in the opposite direction; the magnification is underestimated.  In other lensing simulations, the source position is often fixed and the images are found by an iterative minimization algorithm.  This would give essentially the same result as our nearest point magnification.  Both effects are much smaller for a smaller source size, as they should be.  

Many images were merged because their centroids were within 0.1~arcsec.  In these cases, it makes no sense to take the closest point magnification since the closest point is not unique.  Unsurprisingly, the magnification at the centroid point is an even worse approximation in these cases, as can be seen in figures~\ref{fig:mag_error} and \ref{fig:mag_ratio}.  In exceptional cases, the centroid might not even be in one of the images that are merged.  As expected, these cases only arise when substructure is present.

Figures~\ref{fig:mag_error} and \ref{fig:mag_ratio} should give one pause before using the point source approximation for the magnification in any substructure lensing study or when interpreting the results of any studies that use this approximation.

\subsection{frequency of $\Delta\theta$ - $R_{\rm cusp}$ outliers}
\label{sec:freq-delt-r_rm}

To determine how often it would be expected for a lens to have  $\Delta\theta$ and $R_{\rm cusp}$ values that are inconsistent with a smooth lens model we define a region around the distribution in the case where no substructure is present and find how many simulated lenses lie outside this region when substructure is added.  We define this region by taking bins in $\Delta\theta$ that contain 2000 simulations taking the long-axis and short-axis cases separately.  Upper and lower boundaries within each bin are set such that 2.5\% of the simulations in the bin are greater than the upper bound and an equal number are less than the lower bound.  The bins completely cover the full possible range of $\Delta\theta$.  Without substructure, 5\% of a lens lie outside of this region.  The fraction of simulated lenses outside this region when substructure is added will be called the fraction of outliers.

\begin{figure} \hspace{-0cm} \rotatebox{90}{
\includegraphics[width=6.5cm]{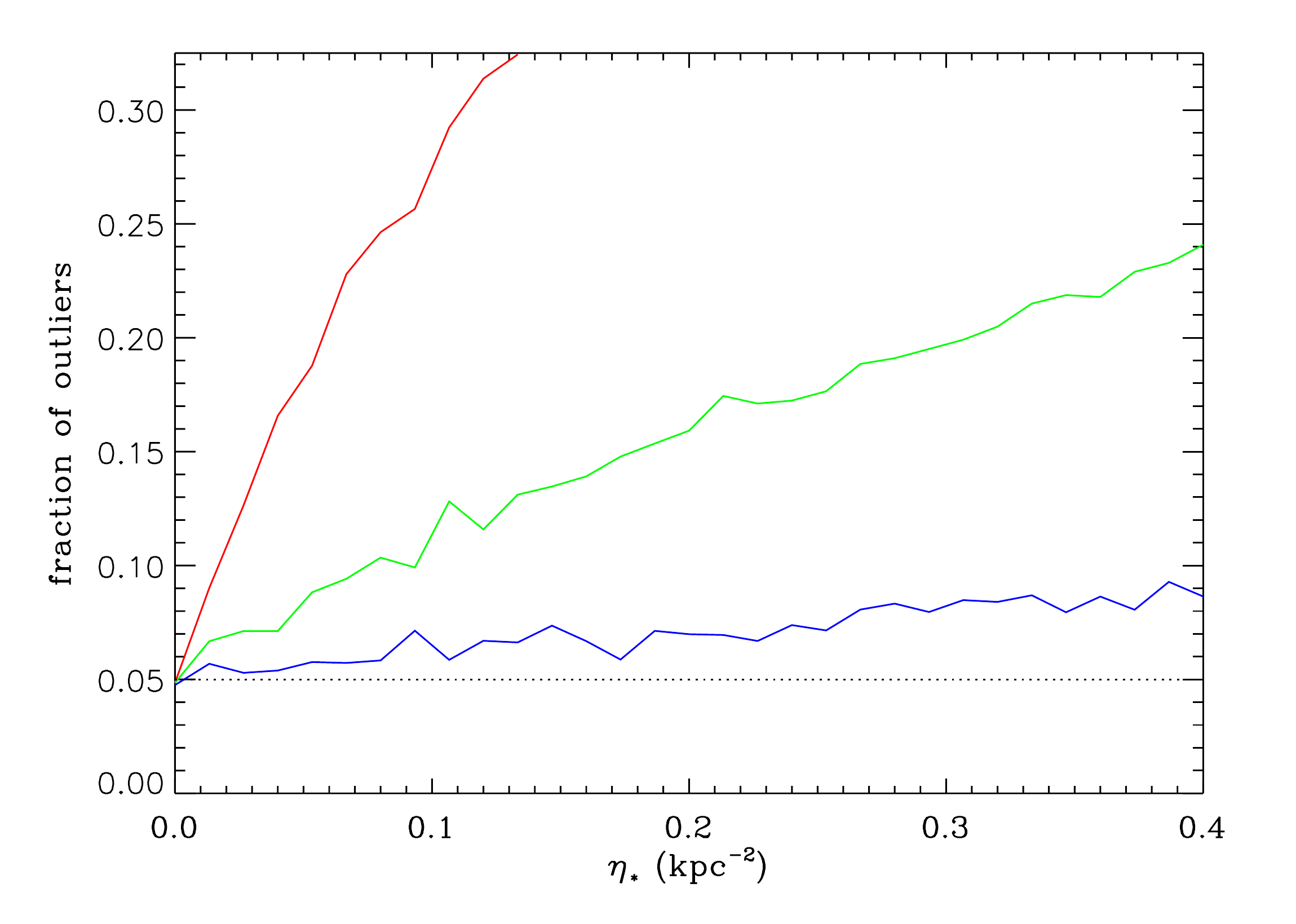}} 
\caption{The fraction of lenses that lie outside the region in $\Delta\theta$-$R_{\rm cusp}$ space contains 95\% of the lenses when there is no substructure (see text for details) as a function of substructure number density.  In all cases $M_{\rm max} = 10^9~\msun$.  The curves are for $M_{\rm min}=10^8\msun$ (red), $M_{\rm min}=10^7\msun$ (green) and $M_{\rm min}=10^6~\msun$ (blue,).  These correspond to sets 1, 2 and 3, respectively from Table \ref{table:simulations}. There are 100,000 simulated lenses used in calculating each point.}
\label{fig:p_eta}
\end{figure}

Figure~\ref{fig:p_eta} shows the fraction of outliers as a function of the substructure surface number density, $\eta_*$, for different substructure minimum masses (simulation sets 1, 2 and 3).  A significant fraction of the lenses are found to be outliers.  The top panels of figures~\ref{fig:pofeta_multi} through \ref{fig:pofeta_multi_3} show the same outlier fraction, but as a function of surface mass density.  

It is surprising that in figure~\ref{fig:pofeta_multi} the outlier fraction appears dependent only on the total surface mass density and not on the lower mass cutoff.  One might think that all the lensing is being done by the most massive substructures and this is why the lower mass cutoff is not important in these cases.   This does not seem to be the case; from set 1 ($M_{\rm min} = 10^8\msun$) to set 3 ($M_{\rm min} =  10^6\msun$) the mass density in the highest decade of mass ($10^8$ to $10^9~\msun$) drops by 60\% for the same total surface mass density and yet the number of outliers is unchanged.

Figure~\ref{fig:pofeta_multi_2} shows the importance of compactness and internal structure on the number of outliers.  The substructure mass function is the same for all the models in this figure.   The slope of the internal density profile, $\beta$, seems to have very little effect on the outlier fraction.  On the other hand, the size of the substructures, or their compactness, does have a strong influence of the outlier fraction.  Between $R_{\rm max} = 0.5~\kpc$ and $R_{\rm max} = 4.0~\kpc$ the fraction decreases significantly.  Since the size--mass relation of the substructures is related to their galactocentric distance through tidal stripping, this sensitivity would provide information on where the substructures are within the lens halo or outside of it.

In figure~\ref{fig:pofeta_multi_3}, the upper substructure mass limit is changed to investigate further the insensitivity to mass range.  It is seen again that for the same mass--size relation the fraction of outliers is dependent on the total surface mass density and relatively insensitive to the upper mass cutoff.  The sensitivity to substructure compactness is again clearly present.  Set 9 with $M_{\rm max}=10^{10}~\msun$ appears to produce slightly less outliers than set 2 with $M_{\rm max}=10^9~\msun$.  This could be because large substructures will sometimes displace the image positions and magnifications significantly while preserving a low $R_{\rm cusp}$ value; the cusp in the caustic is moved, but its shape remains relatively intact.

From the upper panels of figures~\ref{fig:pofeta_multi} through \ref{fig:pofeta_multi_3}, it can be seen that if the size-mass relation is held fixed the outlier fraction is largely a function of the total surface mass density in substructures and not the range of substructure masses.  This conclusion may depend on the function used here ($\alpha = 1.9$).  Further simulations will be needed to investigate this.
Changing the size-mass relation so that the substructures are less dense does reduce the fraction of anomalies (sets 6 and 8).

\section{comparison with data}
\label{sec:comparison-with-data}

\begin{figure} \hspace{-0cm} \rotatebox{90}{
\includegraphics[width=9.5cm]{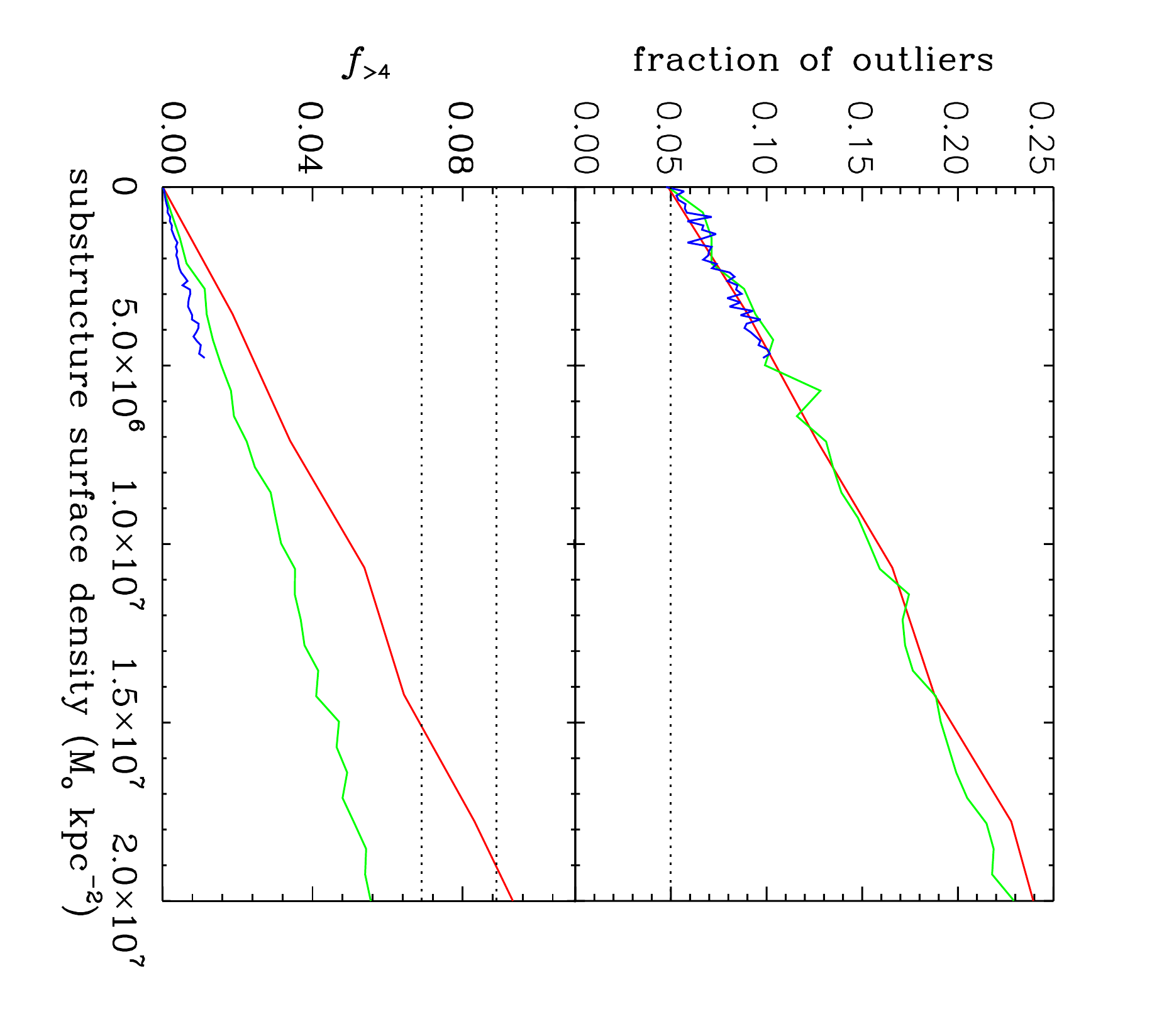}}
\caption{The top panel is the same as in figure~\ref{fig:p_eta}, except it is now as a function of the surface mass density in substructures.  The bottom panel shows the fraction of lenses with more than 4 images with separations of more than 0.1~arcsec and flux ratios of within a factor of 100 (excluding the cases with less than four images).  The colors are the same as in figure~\ref{fig:p_eta}.  The dotted lines in the bottom panel show where there is only a 10\% and 5\% chance of a sample of 32 lenses having no cases of more than 4 images as in the Castles lens sample.} 
\label{fig:pofeta_multi}
\end{figure}

\begin{figure} \rotatebox{90}{
\includegraphics[width=9.5cm]{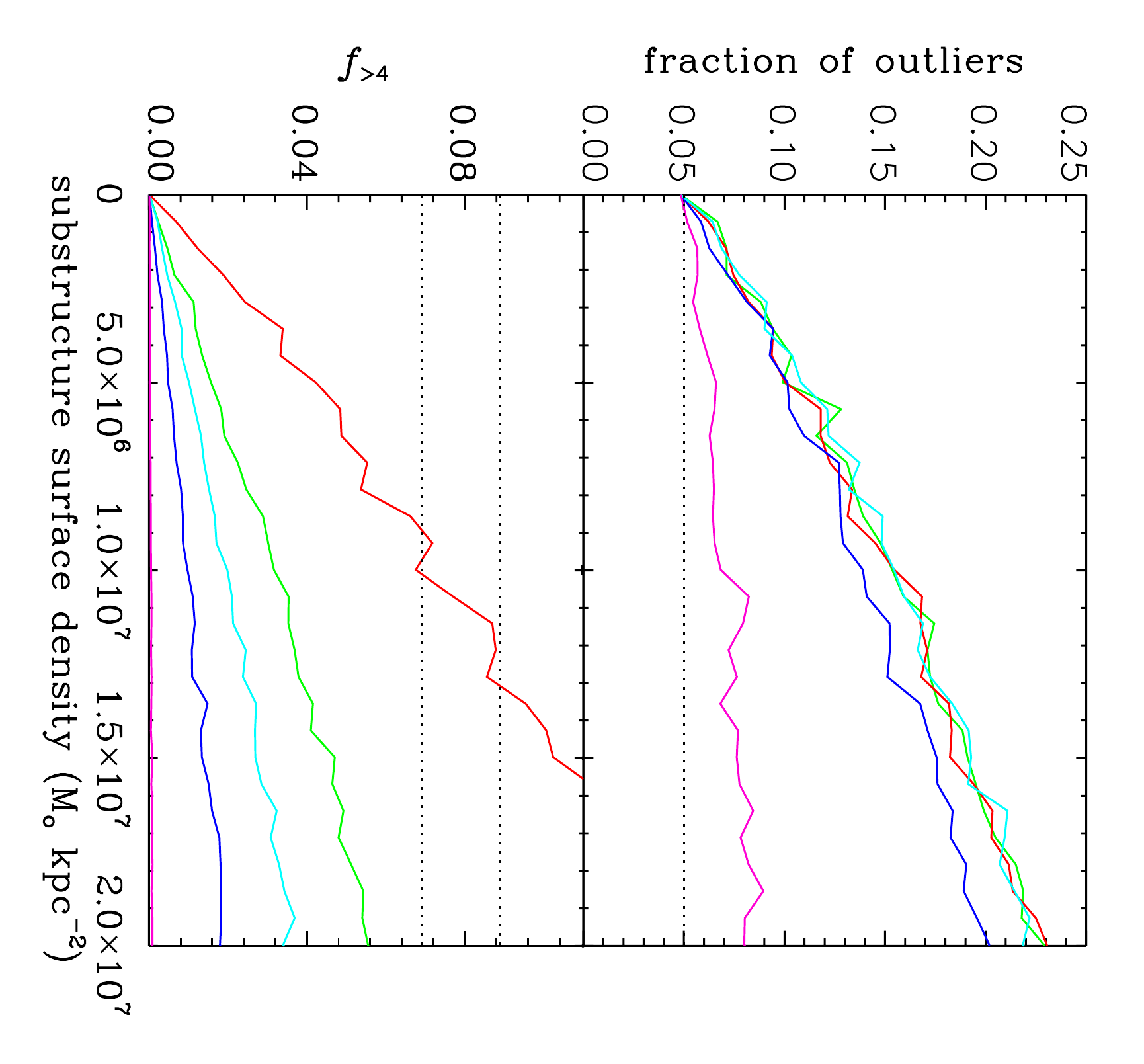}} 
\caption{Same as figure~\ref{fig:pofeta_multi}, but with substructure models with different internal structures.  Green is set 1 (see table~\ref{table:simulations}) as in figure~\ref{fig:pofeta_multi}.  Red is for denser substructures (set 5), while blue and purple are for less dense substructures (sets 6 and 8 respectively).  Cyan is for substructures with less steep mass profiles, but the same sizes as green (set 7).  Changing the internal mass profile has no discernible effect on the frequency of flux anomalies, but has a significant effect on the frequency of high multiplicity lenses.} 
\label{fig:pofeta_multi_2}
\end{figure}

\begin{figure} \rotatebox{90}{
\includegraphics[width=9.5cm]{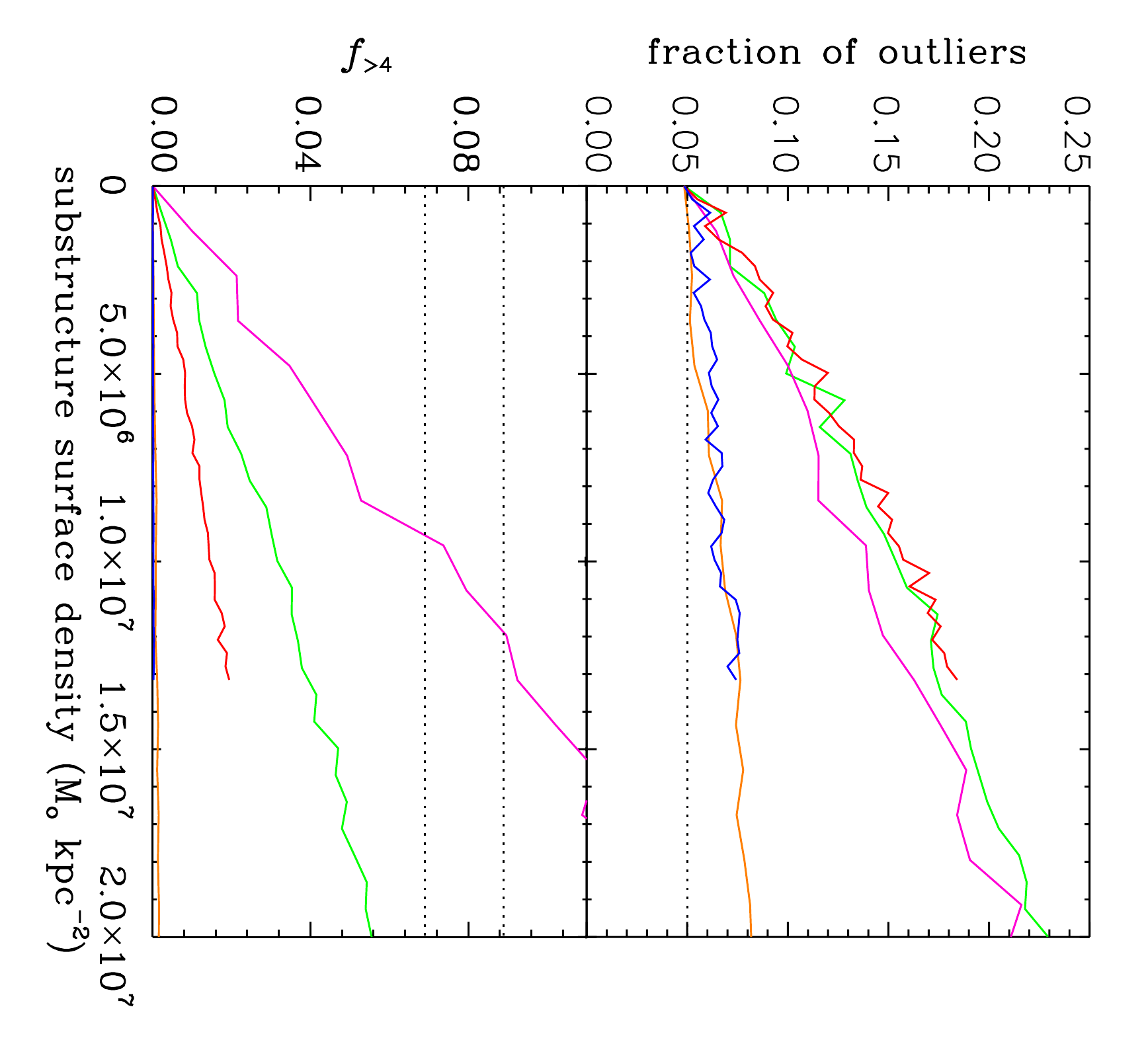}} 
\caption{Same as figure~\ref{fig:pofeta_multi}, but with substructure models meant to explore the importance of the upper mass limit.  Green is set 1 (see table~\ref{table:simulations}) as in figure~\ref{fig:pofeta_multi} and \ref{fig:pofeta_multi_2}. The purple and brown (sets 9 and 10, respectively) have a higher upper mass cutoff of $M_{\rm max}= 10^{10}\msun$ and different mass--size relations.  The red and blue curves (sets 11 and 12) are for a  mass cutoff of  $M_{\rm max}= 10^{8}\msun$.  The $R_{\rm max}$ values are set so that a substructure of the same mass will have the same size in sets 1, 9 and 11. And the same is true for sets 10 and 12.  The compactness clearly has a strong effect.  The fraction of outliers for the green, red and purple are similar indicating that the upper mass cutoff does not have a strong effect on it when the substructures are compact.  In contrast, the number of high multiplicity lenses clearly is dependent on the mass range.} 
\label{fig:pofeta_multi_3}
\end{figure}

To avoid contamination from microlensing by stars in the lens galaxy, differential extinction and variability of the source on time-scales smaller than the image delay times, we compare our simulations only to quad lenses measured in the radio and the mid-infrared.  Since we have not included companion galaxies to the primary lens in our simulations, we also remove lenses with nearby galaxies that appear to have similar masses to the primary.  This removes 1608+656 and 1004+4112 from the list.  There is a very faint dwarf galaxy within the Einstein radius of 2045+265 \citep{2007MNRAS.378..109M}, but we will consider this to be a substructure and not a companion galaxy because it is small.  Lens models show that this substructure would need to be unnaturally elongated to cause the flux anomaly in this system, so there is probably another substructure present.
The lenses must also have a detected lens galaxy which eliminates 0134-0931 and 0128+437.  Table~\ref{table:lenses} lists the lenses used and their $R_{\rm cusp}$ and $\Delta\theta$ values are plotted in figures~\ref{fig:Rcusp1} and \ref{fig:Rcusp2}.

\begin{table}
\caption{Observed lenses used in this analysis}
\begin{tabular}{c|ccrl}
name & band & $\Delta\theta$ & $R_{\rm cusp}$ & reference \\
\hline
\hline
2045+265  & radio & 35.3$^\circ$ & 0.501  & \cite{1999AJ....117..658F} \\
0712+472  & radio & 79.8$^\circ$ & 0.254  & \cite{1998MNRAS.296..483J} \\
1555+375  & radio & 108$^\circ$ & 0.417 & \cite{1999AJ....118..654M} \\
1422+231  & mid-IR & 74.9$^\circ$ & 0.203  & \cite{2005ApJ...627...53C} \\
0414+053  & radio & 101.5$^\circ$ & 0.220 & \cite{1997ApJ...475..512K} \\
2237+030  & radio & 146.3$^\circ$ & 0.357 & \cite{1996AJ....112..897F}\\
1115+080 & mid-IR & 127.5$^\circ$ & -0.043 & \cite{2005ApJ...627...53C}
\end{tabular}
\label{table:lenses}
\end{table}

The most striking thing in figures~\ref{fig:Rcusp1} and \ref{fig:Rcusp2} is that one of the lenses, 2045+265, has significantly higher $R_{\rm cusp}$ than is expected in the absence of substructure, but that all the other lenses have $\Delta\theta$--$R_{\rm cusp}$ values that are not particularly anomalous.  In the bottom two rows of figure~\ref{fig:Rcusp1} it can be seen that if only the low ellipticity lenses (axis ratio $> 0.7$) were considered three or four of the observed lenses would have anomalous $\Delta\theta$--$R_{\rm cusp}$ values.  Since the authors that have compared Nbody simulations to the data using $R_{\rm cusp}$ values in the past have used very few simulated lenses and all with axis ratios $\ge 0.7$ \citep{AMCO04,2006MNRAS.366.1529M,2009MNRAS.398.1235X} it is now not surprising that they concluded that the simulations did not produce enough anomalies.

It should be emphasized that just because the lenses' $\Delta\theta$--$R_{\rm cusp}$ values are not anomalous does not mean that they do not have anomalous flux ratios.  Some of these cases clearly cannot be fit by reasonable models without substructure when all the image positions and fluxes are taken into account \citep{2002ApJ...567L...5M,EW03,2008arXiv0801.2258S}.
With so few observed lenses and only one clear anomaly in $\Delta\theta$--$R_{\rm cusp}$ space, it is impossible to make any strong conclusion about the aloud properties for substructure using only the $\Delta\theta$--$R_{\rm cusp}$ distribution.  About one anomalies out of the seven lenses is about what one would expect from studying the top panels of figures~\ref{fig:pofeta_multi} and \ref{fig:pofeta_multi_2} for a substructure surface density of $\sim 10^7~\msun~\kpc^{-2}$.
Other flux-based constraints are possible and will be investigated in future papers.

We introduce another constraint in the bottom panels of figure~\ref{fig:pofeta_multi} through \ref{fig:pofeta_multi_3} based on the fraction of simulations with more than four images.   (This does not include the central demagnified image that forms near the center of the lens for nonsingular lens mass profiles.  In our case, the mass density in the center of the lens diverges like $\Sigma \propto r^{-1}$, and this image never appears; it is infinitely demagnified.)  Even after merging images with centroids less than 0.1~arcsec apart, there are cases where the substructures cause further splitting of the images.  Of the 32 QSO lenses in the Castles \citep{CASTLES} list of lenses with more than four images and simple lenses, none have more than 4 images of a single source separated by more than 0.1~arcsec\footnote{0134-0931 might have five optical images, but two of them are well within 0.1~arcsec of each other.   1933+503 has 10 radio images, but models show that the best explanation is that there are 3 sources with none of them imaged more than four times \citep{1998MNRAS.301..315N}.  1359+154 does appear to be an honest-to-goodness case of a single QSO with 6 images, but the lens is a group of three galaxies and thus does not pass our no companions cut.}.  This puts a strong constraint on the allowed fraction of lenses that have more than four images, $f_{>4}$.  The probability of getting zero cases of $> 4$ images in 33, given that the probability of getting such a case is $p \simeq f_{>4}$, is a binomial distribution.
There would be less than a 5\% chance of this happening in the observed sample if $f_{>4}$ is greater than 0.089 and less than 10\% chance if $f_{>4} > 0.069$.  These are the dotted lines in the bottom panels of figures~\ref{fig:pofeta_multi} through \ref{fig:pofeta_multi_3}.  For equal surface mass density, more massive substructures cause more high image multiplicity lenses.

The multiplicity constraint does change significantly if the resolution cutoff of $0.1$~arcsec is changed.  There are a large number of lenses where the images are merged in some cases (up to $\sim 20\%$).  With improved resolution or a more careful anaylisis of the data, we believe this constaint could be made significantly stronger.

 Within the ranges of $\eta_*$ studied here, the only models that are limited by this image multiplicity constraint are set 1 (high lower mass cutoff and compact), set 5 (super-compact) and set 9 (high upper mass cutoff and compact).  The constraints are $\Sigma_* < 2.0\times 10^7~\msun~\kpc^{-2} $, $\Sigma_* < 1.2\times 10^7~\msun~\kpc^{-2}$ and $\Sigma_* < 1.2\times 10^7~\msun~\kpc^{-2} $ respectively.  The more compact and massive the substructures are the more high multiplicity cases are created.  This constraint is in contrast to the $R_{\rm cusp}$ constraint which depends only on the mass density and compactness.  With more lenses this constraint could become significantly stronger in the future.

\section{expectations for small-scale structure within the CDM model}
\label{sec:expect-small-scale}

A good point of comparison between lens simulations and Nbody simulations is the fraction of mass in substructure within a projected radius of 10~kpc.  This is easily measured in the simulations and since the Einstein radius is typically around 10~kpc, it is close to what is actually constrained by the lensing data.  In our model, this quantity is given by
\begin{align}
f^{10\rm kpc}_{\rm sub} & \equiv \frac{M_{\rm sub}(R<10~\kpc)}{M_{\rm host}(R< 10~\kpc)} = \frac{G\langle m\rangle \eta_*}{\sigma_*^2} 10~\kpc, \\
& = 1.08\times 10^{-9} ~ \kpc^{2} ~ \Sigma_*,
\end{align}
where the fiducial value $\sigma_*=200~\kms$ has been used.  Note that this fraction scales with host mass in our model and in the simulations.  

\cite{2007ApJ...657..262D}  give $f^{10\rm kpc}_{\rm sub} \simeq 0.003$ for the Via Lactea simulation, and \cite{2009MNRAS.398.1235X} give $f^{10\rm kpc}_{\rm sub} \simeq 0.0025$ with a large scatter in the Aquarius simulations.  These simulations should be resolving substructure to below $10^7~\msun$.  These translate to $\Sigma_* = 2.8\times 10^6~\msun\,\kpc^{-2}$ and $2.3\times 10^6~\msun\,\kpc^{-2}$ respectively.  Accounting for the extra mass below there resolution and judging from figures~\ref{fig:pofeta_multi} through \ref{fig:pofeta_multi_3} we would expect about a 10\% chance of a clear outlier in the $\Delta\theta$--$R_{\rm cusp}$ distribution for the high compactness cases which is consistant with the one out of seven observed.  For the larger size--mass relation (sets 8, 10 and 12) the expected fraction is increased by only a few percent from the no substructure case, but with only one observed outlier, we do not consider this a significant contradition. 

\cite{AMCO04}, \cite{2006MNRAS.366.1529M} and \cite{2009MNRAS.398.1235X} come to the conclusion that the substructure present in the simulations is not enough to cause the observed frequency of $R_{\rm cusp}$ anomalies.  In light of the findings in this paper we believe that these conclusions were flawed because the full range of host lens ellipticiites was not represented in the simulations.    \cite{2006MNRAS.368..599M} may have used too low a substructure mass range ($10^5 - 10^7~\msun$) to cause enough anomalies.

There are a number other complicating factors that make comparing observations to the true predictions of CDM difficult.
For example, the baryons are not accounted for in the Nbody simulations.  This impacts the predictions in several ways.  First, the host galaxy needs to be inserted by hand into these Nbody simulations for them to be realistic lenses.   The mass fraction decreases with the inclusion of baryons.  Second, the baryons are expected to have some effect on the internal structure of the substructures, either expanding or contracting them, which will affect their tidal stripping and disruption in the host halo.  The resident galaxy might also have a significant effect on the survival of substructures.   As discussed in section~\ref{sec:host-lens-model}, the typical galactocentric distance for substructures that are important for lensing is significantly smaller than the typical distance of substructures in general.  The substructure population probed by lensing is likely to be more compact and have a steeper mass function, at least above $\sim 10^8 \msun$, than the general population.  This steepening of the mass function at high masses has been only crudely accounted for in our model by the $M_{\rm max}$ cutoff parameter.  

Because we appear to be consistent with the simulations on the frequency of $R_{\rm cusp}$ anomalies does not mean that some other test, such as fitting each simulated lens to a smooth lens model, would not show some inconsistency.  Modeling the lens puts constraints on the ellipticity.  Our argument is that $R_{\rm cusp}$ is not a good test for the existence of substructure without further constraints.

\section{conclusions \& discussion}
\label{sec:concl--disc}

We have preformed the largest number of lens simulations ever done with finite size sources.  This was made possible by the new adaptive ray-tracing code GLAMER.  We find that accounting for the finite size of the source is necessary for drawing accurate conclusions from the lensed QSO data.

We find rough consistency between the $\Lambda$CDM predictions and observations.  $R_{\rm cusp}$ is found to be a poor discriminator between lenses with substructure and without because of its sensitivity to the ellipticity of the lens.  The distribution of ellipticities used in our lens models is based on the ellipticities of observed lenses so we do no think the ellipticities required to explain the observed $R_{\rm cusp}$ distribution (accepting lens 2045+265) are atypical.  Other methods for comparing observations to models are likely to be more fruitful.  And as the data improves more precise comparisons will be possible.
In addition to the substructure within the primary lens, there should be some contribution from intergalactic small-scale structure \citep{Metcalf04,M04b} so one should expect the limits derived from the data to be somewhat higher than the limits derived from Nbody simulations of individual dark matter halos.  The baryons also clearly play a role in shaping the lensing properties and they are not fully taken into account in the simulations at the necessary resolution.

We have limited our study here to a substructure mass function of the form $dN/dm \propto m^{-\alpha}$ with $\alpha=1.9$.  This seems well motivated by the simulations on small mass-scales, but could be steeper on larger mass-scales because of tidal stripping and disruption in the central regions of the lens. 
With the  $\alpha=1.9$ mass function, the smaller mass substructures plays a smaller part in causing flux anomalies because most of the mass resides in larger mass objects.  This will make it difficult to measure any possible lower mass cutoff using monochromatic QSO lensing alone.  Fortunately there are some other prospects for probing the mass function in the future such as spectroscopic gravitation \citep{MM02} and Einstein rings \citep{2009MNRAS.400.1583V}.  If the slope of the mass function is steeper than $\alpha=1.9$, the smaller structures will play a larger role in the lensing.  

It is clear that what is really required to make a more conclusive measurement of the amount of substructure in dark matter halos is more data.  With 7 lenses, only limited conclusions can be made from a statistical point of view.  We are also vulnerable to systematic errors.  For the kind of study done here, more strong lenses measured in the radio and/or mid-infrared are needed.  Planned large scale imaging surveys\footnote{DES (darkenergysurvey.org) (snap.lbl.gov), LSST (www.lsst.org), PanSTARS (pan-stars.ifa.hawaii.edu),
  VISTA (www.vista.ac.uk), EUCLID (http://www.ias.u-psud.fr/imEuclid/)} expect to increase the number of lensed QSOs in the visible by an order of magnitude so we look forward to  great improvements in this field.

\vspace{0.3cm} 
\leftline{\bf Acknowledgments} 
We would like to thank Simon White and Raul Angulo for useful discussions and comments.  We would like to thank Jacqueline Chen for pointing an important error in an earlier version of this paper and the anonymous referee for very helpful suggestions and criticism.  

RBM's research is part of the project GLENCO, funded under the Seventh Framework Programme, Ideas, Grant Agreement n. 259349.

 \bibliographystyle{/Users/bmetcalf/Work/TeX/apj/apj}
 \bibliography{/Users/bmetcalf/Work/mybib}

\end{document}